# NANOINDENTING THE CHELYABINSK METEORITE TO LEARN ABOUT IMPACT DEFLECTION EFFECTS IN ASTEROIDS

Carles E. Moyano-Cambero[1*], Eva Pellicer[2], Josep M. Trigo-Rodríguez[1], Iwan P. Williams[3], Jürgen Blum[4], Patrick Michel[5], Michael Küppers[6], Marina Martínez-Jiménez[1], Ivan Lloro[1], Jordi Sort[7].

[1]Institute of Space Sciences (IEEC-CSIC), Meteorites, Minor Bodies and Planetary Sciences Group, Campus UAB Bellaterra, c/Can Magrans s/n, 08193 Cerdanyola del Vallès (Barcelona), Spain; moyano@ice.csic.es, trigo@ice.csic.es

[2]Departament de Física, Universitat Autònoma de Barcelona, E-08193 Bellaterra, Spain

[3]School of Physics and Astronomy, Queen Mary, University of London, 317 Mile End Road, E1 4NS London, UK

[4] Institut für Geophysik und extraterrestrische Physik, Technische Universität Braunschweig, Mendelssohnstr. 3, D-38106 Braunschweig, Germany

[5]Lagrange Laboratory, University of Nice, CNRS, Côte d'Azur Observatory, France

[6] European Space Agency, European Space Astronomy Centre, P.O. Box 78, Villanueva de la Cañada E-28691, Spain

[7]Institució Catalana de Recerca i Estudis Avançats (ICREA) and Departament de Física, Universitat Autònoma de Barcelona, E-08193 Bellaterra, Spain

## ABSTRACT

The Chelyabinsk meteorite is a highly shocked, low porosity, ordinary chondrite, probably similar to S- or Q-type asteroids. Therefore, nanoindentation experiments on this meteorite allow us to obtain key data to understand the physical properties of near-Earth asteroids. Tests at different length scales provide information about the local mechanical properties of the minerals forming this meteorite: reduced Young's modulus, hardness, elastic recovery, and fracture toughness. Those tests are also useful to understand the potential to deflect threatening asteroids using a kinetic projectile. We found that the differences in mechanical properties between regions of the meteorite, which increase or reduce the efficiency of impacts, are not a result of compositional differences. A low mean particle size, attributed to repetitive shock, can increase hardness, while low porosity promotes a higher momentum multiplication. Momentum multiplication is the ratio between the change in momentum of a target due to an impact, and the momentum of the projectile, and, therefore higher values imply more efficient impacts. In the Chelyabinsk meteorite, the properties of the light-colored lithology materials facilitate obtaining higher momentum multiplication values, compared to the other regions described for this meteorite. Also, we found a low value of fracture toughness in the shock-melt veins of Chelyabinsk, which would promote the ejection of material after an impact and, therefore, increase the momentum multiplication. These results are relevant considering the growing interest in missions to test asteroid deflection, such as the recent collaboration between the European Space Agency and NASA, known as the Asteroid Impact and Deflection Assessment mission.

**Tables: 2**

**Figures: 4**

# TEXT

## 1. INTRODUCTION

Although the probability of an asteroid causing a catastrophic impact is statistically small (Atkinson et al. 2000), the public concern about impact hazard increased in 2013, when a small asteroid overflew the Russian region of Chelyabinsk, producing a large airburst accompanied by thousands of meteorite specimens falling, with a total mass of ~1000 kg (Nazarov et al. 2013; Ruzicka et al. 2015). The asteroid diameter, estimated at ~18 meters (Brown et al. 2013), was relatively small compared with kilometer-sized bodies capable of producing a mass extinction. Nevertheless, the shockwave released caused significant damage to buildings, and ~1500 people were injured. The Chelyabinsk event shows that even if a highly destructive impact has a very low probability of occurrence, airburst effects can still be dangerous (see e.g. Wasson 2003). Such events can be expected on a once in a decade-to-century scale (see e.g. Atkinson et al. 2000) due to collisions with objects coming directly from the main asteroid belt, or also to disruptive processes that occurred in near-Earth space (Trigo-Rodríguez et al. 2007).

There is an ongoing discussion concerning the best strategy to deal with any potential threatening asteroid (Morrison 2010). Some techniques, such as using a gravity tractor, require years or even decades to be effective (Lu & Love 2005). Kinetic impact strategies, which imply using a projectile to slightly change the orbit of a near-Earth Asteroid (NEA), are technologically more advanced and require a much shorter time scale (Ahrens & Harris 1992). Due to the controversy related to the use of nuclear weapons, non-explosive projectiles are preferred (Koenig & Chyba 2007). Between 2005 and 2007 the European Space Agency (ESA) proposed the Don Quijote mission (Carnelli et al. 2006), with the aim of testing the feasibility of using a kinetic projectile to deflect an asteroid, and also to properly observe and analyze the consequences on the target asteroid. The mission was not adopted, but aspects of it were incorporated in the Asteroid Impact and Deflection Assessment (AIDA) mission. AIDA has been a collaboration between ESA and NASA to develop two complementary spacecraft: the *Asteroid Impact Mission (AIM),* by ESA, and the *Double Asteroid Redirect Test (DART),* by NASA (Michel et al. 2015a, 2016). The two missions are planned to travel to the binary NEA (65803) Didymos, composed of a primary 800 m asteroid and a 150 m satellite. The latter will be impacted by the 300 kg *DART* spacecraft, while *AIM* would characterize the system before and after such event (Michel et al. 2015a). However, AIM did not receive the necessary funding in December, 2016, and, therefore, the future of AIDA is unclear.

The success of AIDA and similar concepts highly depends on the knowledge of the physical (i.e., mechanical) properties of the NEA to be deflected. A proper characterization of these objects is, therefore, required to avoid, or minimize, unexpected outcomes such as a multiple fragmentation (Holsapple & Ryan 2002). Here, we present a laboratory approach using meteorite specimens in order to quantify, in controlled small-scale experiments, mechanical parameters that might be used to predict the effects caused by a projectile on the surface of an asteroid. Nanoindentation is selected here as an almost nondestructive technique, compared to impact tests. The Chelyabinsk meteorite was selected as a good example of the different materials that form small NEAs such as (65803) Didymos. The results obtained here using quasistatic conditions from a Chelyabinsk specimen, are interpreted in view of their correlation with dynamic mechanical parameters that play a role during an impact between the asteroid and an external body.

## 2. RATIONALE FOR SAMPLE SELECTION AND METHODS

Chelyabinsk has been classified as an LL5 and LL6 ordinary chondrite (OC) breccia with an S4 shock stage and exhibiting different lithologies (Bischoff et al. 2013; Kohout et al. 2014; Ruzicka et al. 2015). The most abundant is the light-colored lithology, constituting ~65% of the meteorite. It shows a typical equilibrated chondritic texture, exhibits an intermediate shock state, and contains recrystallized chondrules that are deformed or broken, plus very thin inter-granular metal and troilite veins (Galimov et al. 2013; Kohout et al. 2014; Righter et al. 2015; Ruzicka et al. 2015;). Some LL6 fragments with rare chondrule relicts and highly recrystallized LL5/6 or LL6 regions, with and without shock veins, have been classified as part of the light-colored lithology, but also as different lithologies (Bischoff et al. 2013). The following most common lithology is the dark-colored (or shock-darkened) one, in which only a small fraction of the original equilibrated chondritic texture remains. This lithology contains a much larger amount of inter- and intra-granular thin melt veins of opaque material (metal and troilite) due to shock mobilization (Bischoff et al. 2013; Galimov et al. 2013; Kohout et al. 2014; Reddy et al. 2014; Righter et al. 2015). The light- and dark-colored lithologies are rarely found in the same specimen (Kohout et al. 2014). A third lithology is often encountered together with any of the other two: a fine-grained dark impact-melt lithology containing finely dispersed metal and troilite droplets, variable abundances of mineral and lithic clasts, but no high-pressure phases (Bischoff et al. 2013; Galimov et al. 2013; Kohout et al. 2014; Righter et al. 2015). Due to their similar color, the dark-colored and the impact-melt lithologies have often been considered together as a single lithology (Reddy et al. 2014; Ruzicka et al. 2015). The three lithologies have roughly similar compositions (Galimov et al. 2013; Kohout et al. 2014;), with olivine (~$Fa_{28}$), are strongly affected by shock, and are two to four times more abundant than pyroxene (~$Fs_{23}$) (Galimov et al. 2013; Ruzicka et al. 2015;). Although both minerals are mostly homogeneous in composition (Galimov et al. 2013; Kohout et al. 2014; Righter et al. 2015; Ruzicka et al. 2015), pyroxene is mainly orthopyroxene, in a proportion superior to 2:1 over clinopyroxene (Galimov et al. 2013; Ruzicka et al. 2015). Small anhedral plagioclase (~$Ab_{86}$) grains also show the consequences of shock, and indeed in the dark-colored lithology the isotropy of plagioclase is complete (Galimov et al. 2013; Ruzicka et al. 2015). Opaque minerals consist of 6-7 wt% of troilite and 2-4 wt% of metal phase, the latter being mostly kamacite (~5 wt% of Ni) and taenite (~35 wt% of Ni) (Galimov et al. 2013; Popova et al. 2013; Ruzicka et al. 2015). Chromite, phosphate (apatite), and ilmenite, among others, are accessory minerals (Galimov et al. 2013).

As an LL-type OC, Chelyabinsk can be easily connected with most NEAs, usually associated with S- or Q-class asteroids (Gaffey 1976; Binzel et al. 2001; Vernazza et al. 2008; Reddy et al. 2014). In fact, it has been suggested that ~2/3 of NEAs show a better match with LL chondrites than with the other OC types (Vernazza et al. 2008). It has been found that S- and Q- asteroids probably form as rubble piles (Holsapple 2001) and, therefore, should show a considerable degree of shock and brecciation, similar to what has been seen in the Chelyabinsk meteorite (Bischoff et al. 2013; Ruzicka et al. 2015). Didymos, in particular, has been classified as an Apollo-class asteroid related to the S-complex, and it also has been spectroscopically connected to L/LL-type meteorites (Dunn et al. 2013). As a binary system, the formation of Didymos as a rubble pile is logical (Walsh & Richardson 2006). As well as being a good analog for Didymos, Chelyabinsk also allows for testing the effects of shock on the mechanical properties of OC-forming minerals. Such chondrites originated from the catastrophic disruption of moderately large asteroids, whose fragments formed families with complex collisional histories (Michel et al. 2001, 2015b; Bottke et al. 2015).

We analyzed one of the specimens of Chelyabinsk, the polished thin section shown in Figure 1 (see Appendix). The microstructure of the sample was studied by optical microscopy and scanning electron microscopy (SEM), and the chemical composition of the different regions was determined by energy dispersive X-ray spectroscopy (EDS). The mechanical properties of the specimen were evaluated by nanoindentation (see Appendix) . To extract the local nanomechanical properties of the sample, indentations with a maximum applied force of 20 mN were performed on the different mineral phases comprised in this OC. The results served to test the consistency between our measurements and the ones obtained from previous studies, and as a reference for mechanical properties assessed using higher applied forces. We were also interested in the average mechanical properties of the material composing the Chelyabinsk meteorite as a whole, as they are important to model the eventual response of NEAs to impacts. Thus, we also performed larger indentations, with an applied force of 500 mN. With such configuration we get rid of the indentation size effect, which is a progressive increase of hardness observed for low indentation forces and ascribed to a variety of factors (see, e.g. Nix & Gao 1998; Gerberich et al. 2002). Thus, the results obtained from indentations performed using 500 mN should be more representative of the real behavior. An array of 16 large indentations was performed on each of the lithologies and regions identified on our selected specimen (Figure 1): the light-colored and the impact-melt lithologies, and a thick black shock-melt vein consisting of a fine-grained silicate matrix with abundant metal and troilite inside (Bischoff et al. 2013; Ruzicka et al. 2015). In order to calculate their mean mechanical properties, we averaged the results of the large indentations performed on the silicates of each region (olivine, pyroxene, and plagioclase), as these minerals account for ~90% of the material in Chelyabinsk (Galimov et al. 2013). We believe that this combination of several indentations can provide a good estimation of the average mechanical properties of each region. Although higher applied loads could result in even more representative values, larger forces cannot be applied with our indenter. The hardness ($H$) and reduced Young's modulus ($E_r$) values were determined using the conventional method (Oliver & Pharr 1992; see Appendix). The elastic recovery was evaluated as the ratio between the elastic and the total (plastic + elastic) energies during nanoindentation, $W_{el}/W_{tot}$ (see Appendix).

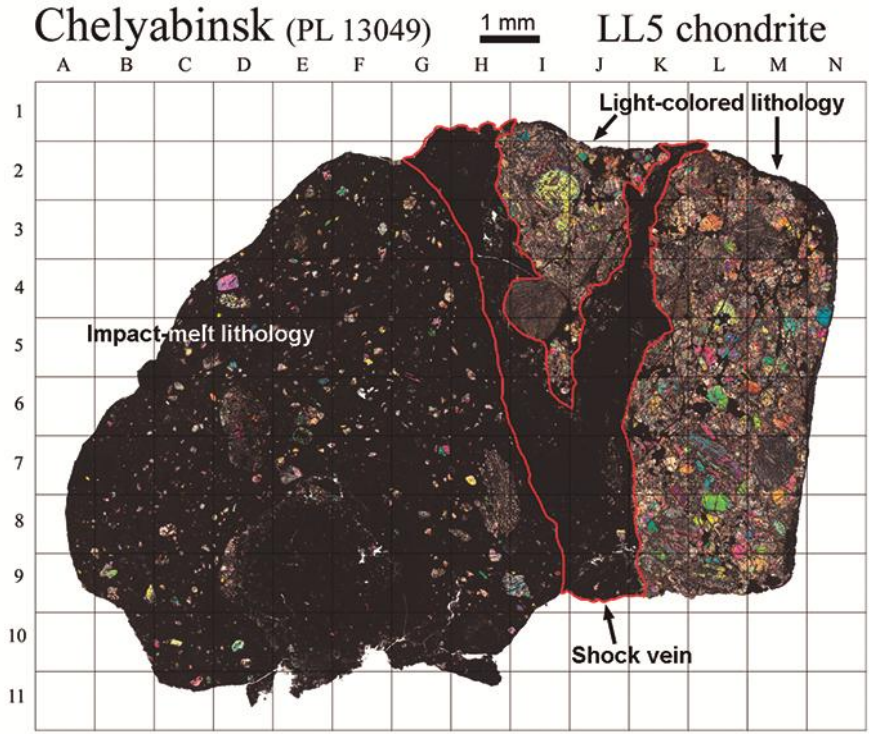

**Figure 1. Transmitted light optical mosaic from the Chelyabinsk PL 13049 thin section.** The three different regions studied here are highlighted. The size of each square of the grid is 1 mm$^2$.

## 3. RESULTS

A summary of the calculated mechanical properties of the main minerals obtained after local indentations (maximum load of 20 mN) is presented in Table 1, and representative load-displacement curves acquired from each mineral phase are shown in Figure 2. According to our results, olivine shows higher hardness and reduced Young's modulus than pyroxene, while both properties notably decrease for plagioclase. It has to be taken into account that the indented olivine included inter-granular metal and troilite veins, which probably increase the variability in mechanical properties. Lower average values of both mechanical parameters are obtained for troilite. For metal grains, composed of kamacite and taenite in variable proportion, rather low hardness and relatively low reduced Young's modulus were found. Taenite has a lower reduced Young's modulus and similar (but slightly lower) hardness than kamacite. The last mineral phase indented was chromite, which shows high values of both $H$ and $E_r$. Concerning the elastic recovery, the highest mean values are achieved in regions where pyroxene and plagioclase are mixed. Chromite also shows a high elastic recovery, whereas troilite and metal inclusions show much lower values.

**Table 1**

Average mechanical properties of Chelyabinsk minerals

| Mineral Phase | $E_r$ (GPa) | $H$ (GPa) | $W_{el}/W_{tot}$ |
|---|---|---|---|

| Mineral | $E_r$ | $H$ | $W_{el}/W_{tot}$ |
|---|---|---|---|
| Olivine | 136 ± 5 | 13.6 ± 0.9 | 0.551 ± 0.023 |
| Pyroxene | 122 ± 11 | 11.9 ± 2.2 | 0.59 ± 0.03 |
| Pyroxene + Plagioclase | 71 ± 5 | 9.6 ± 1.0 | 0.720 ± 0.025 |
| Troilite | 71 ± 8 | 5.1 ± 0.8 | 0.45 ± 0.05 |
| Taenite | 82 ± 6 | 3.05 ± 0.29 | 0.232 ± 0.007 |
| Kamacite | 127 ± 16 | 3.58 ± 0.24 | 0.20 ± 0.06 |
| Chromite | 131 ± 3 | 15.9 ± 1.3 | 0.666 ± 0.016 |

**Notes.** Reduced Young's modulus ($E_r$), hardness ($H$) and elastic recovery ($W_{el}/W_{tot}$) of the same mineral phases from where the curves at Figure 2 were obtained. Each was calculated by averaging the results obtained from several small indentations (up to 20 mN).

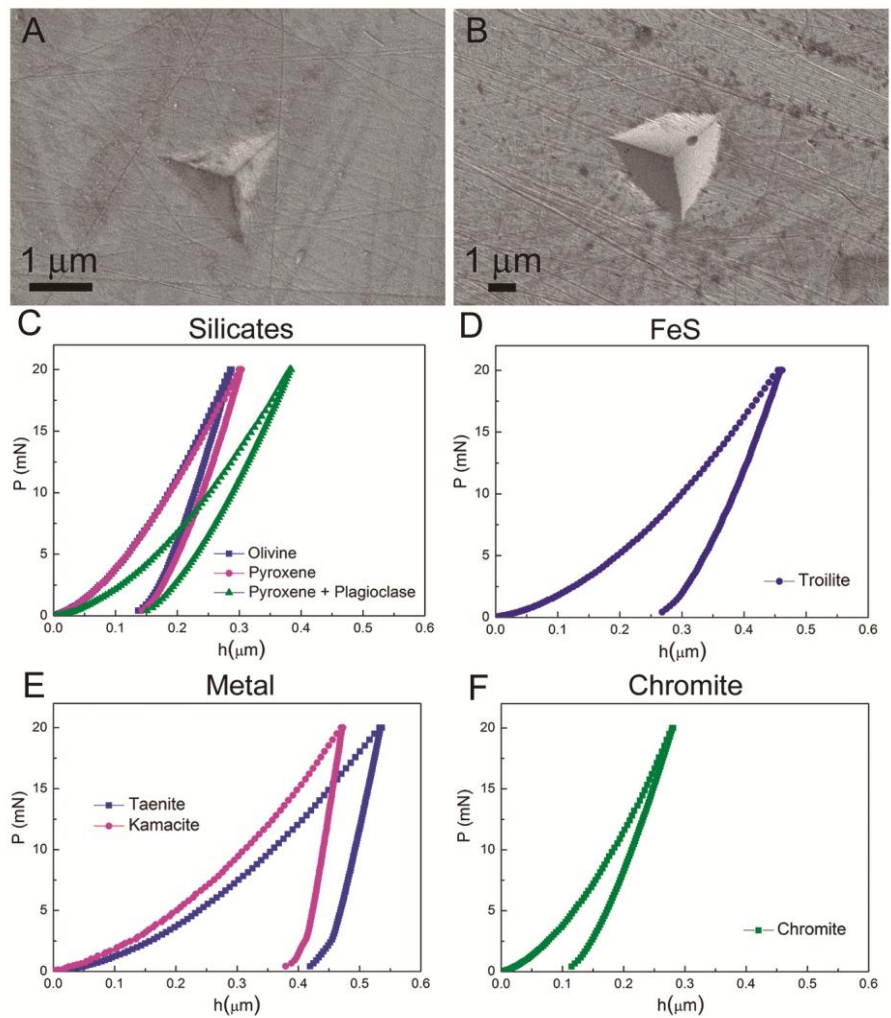

**Figure 2. Low-load indentations (up to 20 mN) performed on the Chelyabinsk meteorite.** At the top, back-scattered electron (BSE) images show indentations on: olivine (**A**), and metal (**B**, taenite or kamacite). Indentation curves for different mineral

phases are shown below. Silicates (**C**): Olivine (Fo$_{75}$, plus tiny troilite veins), Pyroxene (En$_{80}$), and pyroxene + plagioclase (pyroxene plus a small amount of Ab$_{90}$ plagioclase). Troilite (**D**): contains ~54% of atomic S. Metal grains (**E**): taenite (~35% of atomic Ni) and kamacite (~5% of atomic Ni, plus a small amount of troilite). The chromite curve is shown in (**F**). The selected curves are considered representative of the mean mechanical properties of each mineral phase (shown in Table 1).

The results obtained after performing larger indentations (maximum load of 500 mN) to obtain combined mechanical properties are summarized in Table 2, while representative indentation curves for each lithology or region are shown in Figure 3. The impact-melt lithology shows the highest hardness, while the highest values of reduced Young's modulus correspond to the black shock-melt vein. The three regions show very similar values of elastic recovery.

**Table 2**

Average mechanical properties of Chelyabinsk regions

| Region | $E_r$ (GPa) | $H$ (GPa) | $W_{el}/W_{tot}$ |
|---|---|---|---|
| Light-colored lithology | 69 ± 8 | 9.7 ± 2.1 | 0.65 ± 0.07 |
| Impact-melt lithology | 71 ± 8 | 12.2 ± 2.2 | 0.69 ± 0.05 |
| Shock-melt vein | 77 ± 8 | 11.8 ± 2.3 | 0.679 ± 0.027 |

**Notes.** Reduced Young's modulus ($E_r$), hardness ($H$) and elastic recovery ($W_{el}/W_{tot}$) were calculated averaging the results obtained from several large indentations (up to 500 mN).

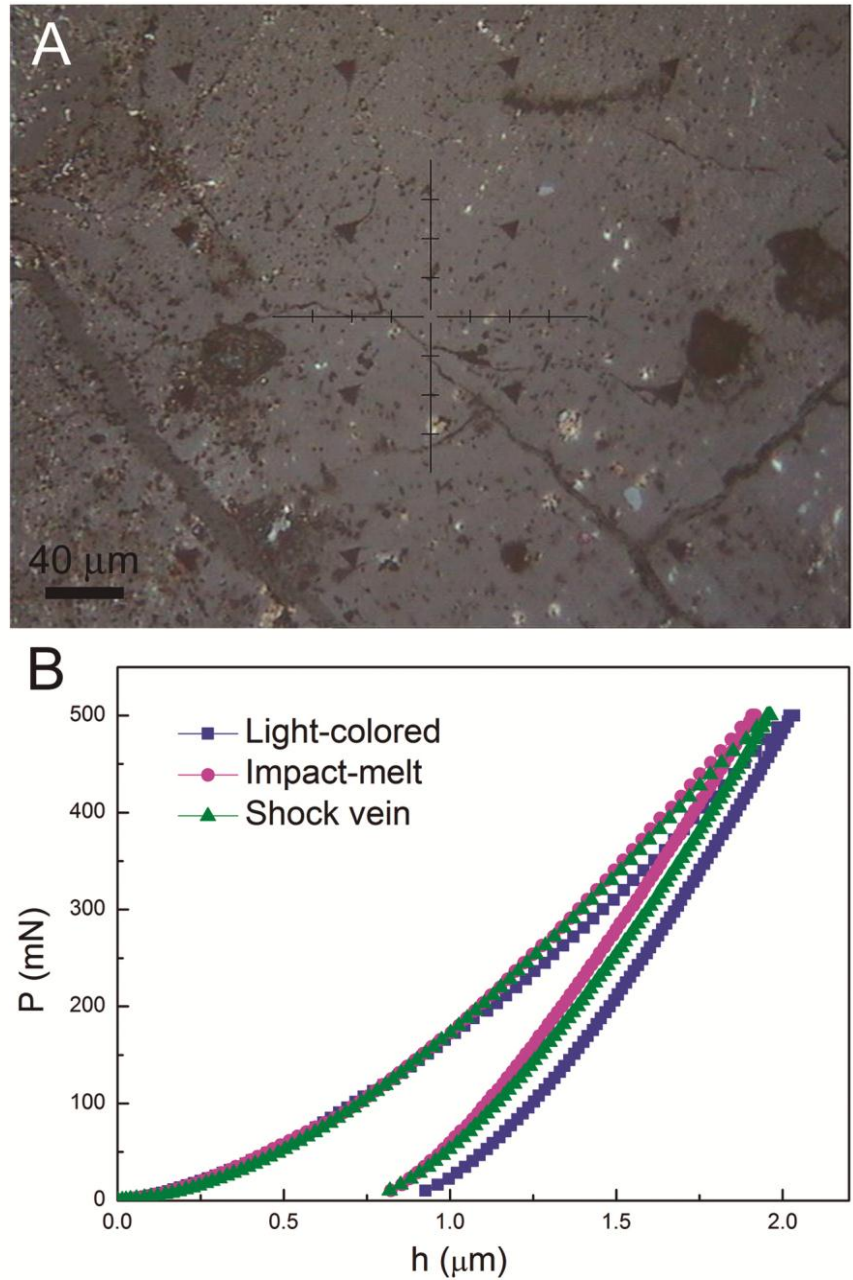

**Figure 3. High-load indentations (up to 500 mN) on the Chelyabinsk lithologies.** At the top, an optical image (**A**) shows an array of 16 high-load indentations on the shock-melt vein. Below (**B**), three indentation curves representative of the mean mechanical properties obtained from the three regions analyzed with the high-load indentations (see data in Table 2).

Remarkably, small cracks are sometimes formed at the edges of the indentations performed using 500 mN load. These cracks are the result of localized fracture, and are almost only observed in the shock-melt vein, as can be seen in Figure 4. The formation and length of these indentation cracks can be correlated with fracture toughness of the indented regions, with longer cracks being indicative of materials more prone to fragmentation (see Appendix). The length of the most significant cracks we could find on our indentations is between 7 and 13 µm, from the center of the indentation to the end of the crack. Thus, with an average $c$ (length from the center of the indentation to the end of the crack) of ~10 µm, we obtain a fracture toughness of 0.62 ± 0.12 MPa·m$^{1/2}$.

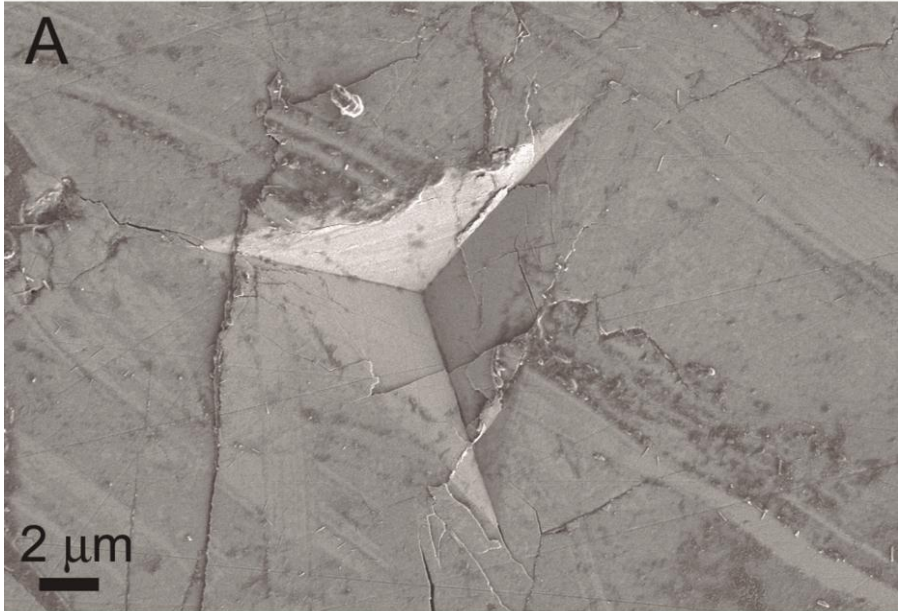
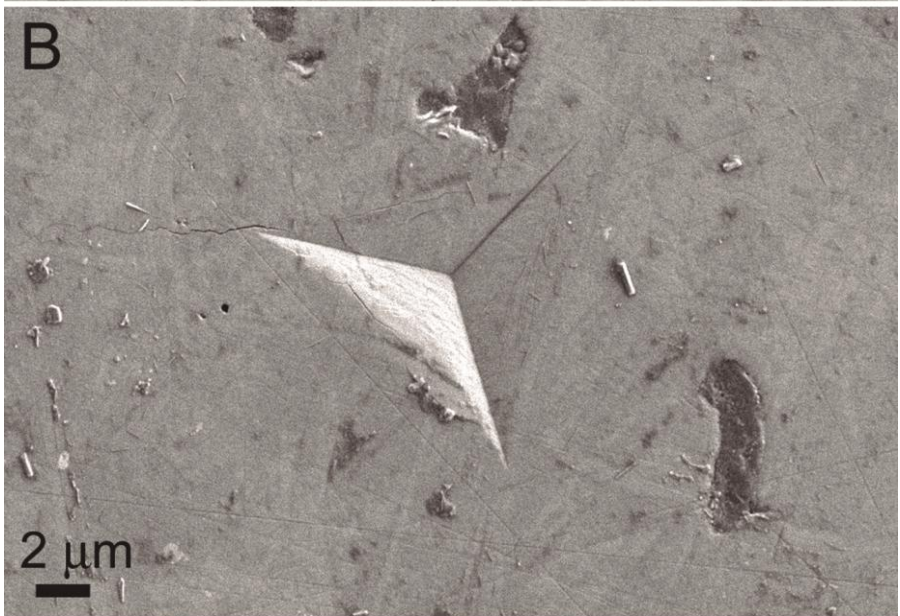
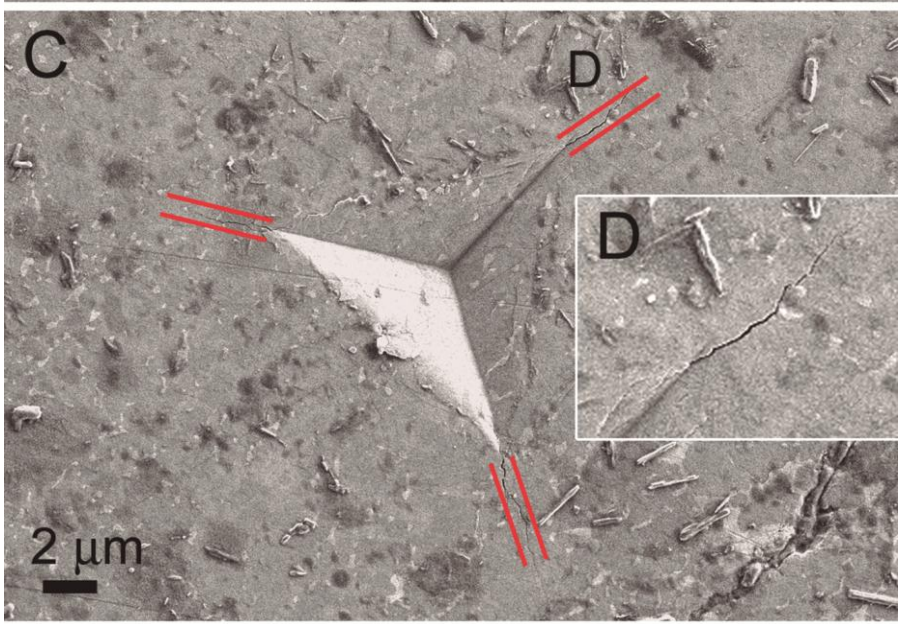

**Figure 4. Fractures after high-load (up to 500 mN) indentations on the three regions analyzed on this Chelyabinsk specimen.** Three BSE images show high-load indentations on: (**A**) the light-colored lithology; (**B**) the impact-melt lithology; (**C**) the shock-melt vein. Fractures can only be clearly identified in the shock-melt vein (surrounded by parallel red lines in (**C**), and shown in more detail in (**D**).

The reduced Young's modulus can be easily related with the Young's modulus (*E*) if one knows the Poisson's ratio (*v*) of the indented material (see Appendix). Using the Poisson's ratio defined for OCs in previous studies (see e.g. Yomogida & Matsui 1983), we see that the reduced Young's modulus values obtained here for the Chelyabinsk meteorite are consistent with the Young's modulus measured previously for ordinary chondrites, between 10 and 140 GPa (see e.g. Yomogida & Matsui 1983; Flynn 2005). As expected, our results indicate that Chelyabinsk has a much higher average reduced Young's modulus than the reported for carbonaceous chondrites, typically around 20 GPa (Britt et al. 2016). Concerning hardness, there is a dearth of previous studies where nanoindentation is applied to meteorites, hence preventing us from an extensive comparison with the literature. However, our measurements of hardness in kamacite and taenite seem to be as expected (Brusnitsyna et al. 2016), thus making our results reliable. Similar to reduced Young's modulus, the nanoindentation hardness of carbonaceous chondrites must be orders of magnitude lower than that of Chelyabinsk (see Appendix), mainly due to the high porosity of carbonaceous chondrites, which can exceed 30% (Consolmagno et al. 2008; Macke et al. 2011; Pellicer et al. 2012). OCs typically show porosities around 5%-10% (Consolmagno et al. 2008), and indeed the values reported for Chelyabinsk range between 2% and 11%, with an average value of ~6%. However, the porosity is almost identical all around Chelyabinsk (Kohout et al. 2014), and, therefore, changes in porosity cannot be considered the cause for the measured differences in mechanical properties between the different investigated regions.

## 4. DISCUSSION

As an LL5-6 OC breccia, Chelyabinsk is representative of the properties exhibited by the surface of a heterogeneous asteroid with distinguishable lithologies, at least at mm size scale as exemplified by our sample. The two lithologies and the shock-melt vein studied in this work are rather similar in chemical composition, mineralogy, and porosity (Galimov et al. 2013; Kohout et al. 2014). Therefore, the differences in mechanical properties between them have to be ascribed to other factors. Also, it does not seem that the hardness values obtained are strongly affected by the value of the applied indentation load (Nix & Gao 1998), since the values of *H* at 20 mN and 500 mN are not significantly different. This could mean that these loads are already high enough to avoid the apparent hardening phenomena related to the indentation size effects. However, our results reveal that when jumping from the in-situ properties of mineral phases (Table 1, Figure 2) to the overall behavior of lithologies or shock-melt veins (Table 2, Figure 3), there is an evident decrease in reduced Young's modulus. Porosity can explain this decrease with applied indentation load (Asmani et al. 2001). As the load is increased, the influence of porosity is exacerbated since the probability to encounter voids in the sample increases. For the same reason, hardness obtained with larger load indentations would probably be smaller than what is shown here (Palchik & Hatzor 2004). The presence of various minor phases and their interaction with the matrix, mainly the melt veins filled with both troilite and metal, possibly alter the resulting mechanical properties of the lithologies, too. In this sense, indentations performed using 500 mN are more representative of the real behavior than those carried out using 20 mN, and are, therefore, better to calculate average bulk properties of the lithologies, as expected. In turn, repetitive shock in the impact-melt lithology and the shock-melt vein can induce a refinement of the mean particle size while also increasing the amount of structural defects (e.g., dislocations), hence leading to mechanical hardening with respect to the light-colored lithology, due to the Hall-Petch strengthening relationship (Gil Sevillano et al. 1980). Finally, it should be noted that the extrapolation of material properties to length scales much larger than the micrometer-sized regions sampled by the 500 mN indentations should be made with caution. It is well known that the strength of bulk materials consisting of multiple phases with dissimilar properties can be very nonlinear (see e.g., Tullis et al. 1991, Durham et al. 2009). The strain rate flow law for an aggregate does not necessarily follow a simple power law, as for monomineralic aggregates. However, here, the mechanically hard phases (olivine, pyroxene) constitute the majority phases in the meteorite and, therefore, the polyphase aggregate can be considered, in a first approximation, as a load-bearing strong framework with a relatively small volume proportion of weaker phases (troilite, taenite, kamacite). In such a case (i.e., with no obvious connection between the weak inclusions), deformation of the aggregate involves strain of the stronger matrix, while not much concentration of strain can occur in the inclusions. This is possibly why the hardness of the different lithologies (with values ranging between 9.7 GPa and 12.2 GPa, Table 2) is close to that of the mechanically hard phases (Table 1), since a load-bearing framework of the stronger material is formed. Otherwise the behavior of the aggregate bulk material would be similar to that of the weak phase alone, as reported by Tullis et al. (1991).

In order to take into account these results in impact scenarios, the key parameter to be considered is the $\beta$ parameter, known as the momentum multiplication factor (see Appendix). If $\beta > 1$ there are fragments ejected after the impact and the impact itself becomes more efficient, due to the "momentum multiplication effect" (Hoerth et al. 2015). Several models have tried to account for the effects of collision and impact on the momentum multiplication in brittle, porous materials (Benz & Jutzi 2006; Hoerth et al. 2015), and in all these models material parameters play a key role. Momentum multiplication can be assessed considering its dependence on porosity and strength of the target (Hoerth et al. 2015), besides from the influence of impact velocities and densities (or masses) of the colliding objects, their size, and other properties (see e.g. Schultz 1993; Holsapple & Housen 2012; Jutzi & Michel 2014; among many others). In solids with low porosity, like the Chelyabinsk meteorite, momentum multiplication is more pronounced, since material ejection is then more directional and no impact energy is dissipated in the form of pore compaction (Hoerth et al. 2015). Then, the role of strength can be considered using the strength-dominated cratering model (see Appendix). According to this model, a lower porosity increases the momentum multiplication, and hence an impact on an asteroid like Chelyabinsk would be more effective in deflecting its trajectory than on NEAs with higher porosity. Also, for a given porosity, meteorites with lower hardness (and, therefore, lower strength) will lead to larger values of $\beta$ and higher efficiency of the impact, since the formation of the impact crater is promoted (see Appendix). Different definitions of strength (compressive strength, yield strength, ultimate strength, etc.) can be related to each other on predictable ways, and, therefore, can be used to estimate the $\beta$ parameter (Holsapple 2009). We chose the compressive yield stress ($\sigma_C$), because it can be related with hardness through the expression $H = C\sigma_C$ (see Appendix). Following the typical definition of the constraint factor $C$ (with a value around 3 for metals), the values of $\sigma_C$ that can be inferred from or measurements of $H$ would be much higher than the compressive strengths found in some previous studies (Buddhue 1945; Kimberley & Ramesh 2011), which are of the order of 10 to a few hundred MPa. Although $\sigma_C$ and compressive strength are not the same, they should be fairly similar. The apparent discrepancy with previous works comes from the specific correlation between $H$ and $\sigma_C$, which varies for different materials, scales, and techniques (see Appendix). Indeed, $C$ can attain values much higher than 3 for ceramics and brittle materials (Zhang et al. 2011). Since $C$ is difficult to be determined, we calculated the ratio between $\beta$-1 from the light and the impact-melt lithologies, instead of their $\beta$ parameter (see Appendix). Assuming a scaling parameter ($\mu$) between 0.4 and 0.55, we see that the $\beta$-1 of the light-colored lithology of Chelyabinsk is between 5% and 20% higher than for the impact-melt lithology. For Chelyabinsk-like asteroids this implies that an impact in objects enriched in light-colored lithology would be more efficient than in others where the impact-melt lithology predominates. The light-colored lithology can be easily distinguished spectroscopically from the dark-colored and the impact-melt lithologies, due to the remarkable darkening of the latter two (Popova et al. 2013; Reddy et al. 2014). This would allow an asteroid impact mission to select the most suitable region to be targeted in order to obtain a more efficient deflection.

We have also seen that although the reduced Young's modulus of the two lithologies and the shock-melt vein are rather similar, the prominent formation of cracks after indentations on the latter is indicative of the ease to create fractures within these veins. It is likely that the amount of mass ejected after an impact depends on fracture toughness, with low fracture toughness values promoting larger ejecta mass, and, therefore, higher momentum multiplication (Walker & Chocron 2015). Not many studies calculate the fracture toughness of meteorites, but our result of $0.62 \pm 0.12$ MPa·m$^{1/2}$ is clearly lower than the 2 MPa·m$^{1/2}$ estimated by some other authors (Walker & Chocron 2015). This is indicative of the ease with which these shock-melt veins can be broken, which would promote the fracturation and consequent ejection of surface materials, promoting the momentum multiplication.

In order to use these results in the frame of an impact deflection mission such as the AIDA, it is necessary to understand the connection between quasistatic indentations and dynamic indentations or impacts. Although it has been observed that for most brittle materials dynamic hardness values are larger than quasistatic values, typically by 10-25% (Anton & Subhash 2000; Wheeler 2009), important mechanical behavior properties still hold for the case of dynamic tests (see Appendix). The difference in scale between a centimeter-sized sample and an asteroid can also have an important effect on the effective mechanical properties. Although several studies have already considered how hardness, strength, and momentum multiplication are affected by variations in size (Schultz 1993; Holsapple & Housen 2012; Jutzi & Michel 2014), a deeper understanding provided by real-scale experiments is required for a proper extrapolation of properties from meteorites to asteroids.

## 5. CONCLUSIONS

Using the nanoindentation technique, we have studied the mechanical properties of one thin sample of the Chelyabinsk meteorite. Since this meteorite is an LL5-6 OC breccia, it is a good proxy for the surface materials of heterogeneous NEAs with distinguishable lithologies and some of the most abundant chondritic materials impacting the Earth. Its study provides constraints to improve our understanding of the mechanical response of such bodies. We summarize our conclusions as following:

1. The value of the applied indentation load (20 mN or 500 mN) does not significantly affect hardness, but reduced Young's modulus decreases notably when moving from the in-situ properties of mineral phases to the overall behavior of lithologies or shock-melt veins. Porosity can explain this decrease, since with larger loads the probability to encounter voids in the sample increases.
2. The differences in mechanical properties between the two lithologies and the shock-melt vein studied in this work cannot be attributed to variations in chemical composition, but the presence of minor phases and melt veins can possibly affect them. Also, lower mean particle size produced by repetitive shock implies increasing the structural defects and, therefore, the mechanical hardening, which increases the hardness of the impact-melt lithology and the shock-melt.
3. Indentations produce cracks in the shock-melt vein, providing as a result a low fracture toughness value, which is indicative of the ease to create fractures within these veins. Low fracture toughness can promote the ejection of surface materials after an impact, therefore increasing momentum multiplication. The shock-melt veins are, therefore, one important structural weakness of Chelyabinsk-like asteroids.
4. Since for a given porosity lower hardness implies larger momentum multiplication, asteroids dominated by light-colored lithology would be easier to deflect than asteroids mainly composed of the impact-melt lithology. As they can be easily distinguished spectroscopically, an asteroid impact mission would be able to select the region where an impact would be more efficient.
5. Our results represent a first step in the use of nanoindentation as a technique to acquire additional insight into the mechanical properties of chondritic bodies, and support AIDA and other future asteroid deflection missions to palliate unexpected impact hazard on human beings.

**Acknowledgements:**

CEMC, JMTR, and MMJ acknowledge funding support from AYA 2015-67175-P. This work has been partially funded by the 2014-SGR-1015 project from the Generalitat de Catalunya and the MAT2014-57960-C3-1-R project from the Spanish Ministerio de Economía y Competitividad (MINECO), cofinanced by the 'Fondo Europeo de Desarrollo Regional' (FEDER). Dr. Eva Pellicer is grateful to MINECO for the "Ramon y Cajal" contracts (RYC-2012-10839). Jürgen Blum acknowledges support by the Deutsche Forschungsgemeinschaft in the framework of FOR2285 under grant BL 298/24-1. Addi Bischoff kindly provided the Chelyabinsk thin section used for this study. This study was done in the frame of a PhD. in Physics at the Autonomous University of Barcelona (UAB) under the direction of JMTR (IEEC-CSIC).

# APPENDIX

## A.1. Technical Specifications

In this study we used a thin (~30 µm) section (by the name of PL 13049) from one specimen of the Chelyabinsk meteorite, kindly provided by Addi Bischoff. The section was polished to mirror-like appearance using diamond paste.

Two high-resolution mosaics of the section were created from separate 50X images taken with a Zeiss Scope Axio petrographic microscope. They were composed of reflected and transmitted light images (Figure 1), respectively. Those mosaics allowed us to establish target features and regions to be analyzed with scanning electron microscopy (SEM), Energy Dispersive X-ray spectroscopy (EDS) and nanoindentation. A 1 mm$^2$ grid was superimposed to locate and naming the different features under study through the specimen (Figure 1).

SEM allowed us to study the microstructure of the specimen, while the chemical composition of the different regions was determined with EDS. SEM images were taken on a Zeiss Merlin field emission (FE) SEM at 1.20 kV. The same instrument allowed acquiring EDS patterns at 15 kV.

## A.2. Calculations Using Nanoindentation Data

Nanoindentation consists in applying a controlled load into a sample through the use of a hard indenter. The indenter pushes the surface while increasing load, up to a specific maximum load. The maximum depth achieved is then measured. When the indenter is unloaded, the sample surface pushes back due to elasticity. The obtained load-displacement curves provide information about the deformation mechanisms (elastic, and plastic), and the elastic recovery, through the loading and unloading curves, respectively. The hardness ($H$) and reduced Young's modulus ($E_r$) values are determined from these curves using the method of (Oliver & Pharr 1992). From the initial unloading slope, the contact stiffness, $S$, are determined as:

$$S = \frac{dP}{dh} \quad (1)$$

where $P$ and $h$ denote the applied force and the penetration depth during nanoindentation, respectively. The reduced Young's modulus is evaluated based on its relationship with the contact area, $A$, and the contact stiffness:

$$S = \beta' \frac{2}{\sqrt{\pi}} E_r \sqrt{A} \quad (2)$$

Here, $\beta'$ is a constant that depends on the geometry of the indenter ($\beta' = 1.034$ for a Berkovich indenter according to Fischer-Cripps 2004). $E_r$ is defined as follows:

$$\frac{1}{E_r} = \frac{1-v^2}{E} + \frac{1-v_i^2}{E_i} \quad (3)$$

The reduced Young's modulus takes into account the elastic displacements that occur in both the specimen, with Young's modulus $E$ and Poisson's ratio $v$, and the diamond indenter, with elastic constants $E_i$ and $v_i$. Note that for diamond, $E_i = 1140$ GPa and $v_i = 0.07$. Hardness can be calculated using the following expression:

$$H = \frac{P_{Max}}{A} \quad (4)$$

where $P_{Max}$ is the maximum force applied during nanoindentation. The elastic recovery is evaluated as the ratio between the elastic and the total (plastic + elastic) energies during nanoindentation, $W_{el}/W_{tot}$. These energies are calculated from the nanoindentation experiments as the areas between the unloading curve and the displacement axis ($W_{el}$) and between the loading curve and x-axis ($W_{tot}$) (Fischer-Cripps 2004).

The nanoindenter used here was a UMIS equipment from Fischer-Cripps Laboratories, operated in the load control mode and using a Berkovich pyramidal-shaped diamond tip. The maximum load applied was of 20 mN for low-load local indentations, and up to 500 mN for larger indentations (the maximum available force for our indenter). All indentations reached depths between 0.2 and 2 µm (see Figures 2 and 3), but the compressive stresses caused by the indenter are not limited to the size of the indent. Therefore, the depth affected and measured by the nanoindentations can go down to ~5 µm for 20 mN loads, and to 20 µm for the 500 mN loads, not deep enough to be disturbed by the properties of the glass supporting the sample (Fischer-Cripps 2004). From the load displacement curves we obtained the hardness ($H$), reduced Young's modulus ($E_r$) and elastic recovery ($W_{el}/W_{tot}$) of the sample with the method described above (Oliver & Pharr 1992). The thermal drift during nanoindentation was kept below 0.05 nm s$^{-1}$. Proper corrections for the contact area (calibrated with a fused quartz specimen), initial penetration depth and instrument compliance were applied.

Whenever cracks are formed after indentations, the formation and length of these cracks can be correlated with fracture toughness of the indented regions. For a Berkovich indentation impression, the fracture toughness, $K$, can be given as (Fischer-Cripps 2004):

$$K = k\left(\frac{E}{H}\right)^n \frac{P}{c^{3/2}} \tag{5}$$

where $k$ is an empirical constant close to 0.016, $P$ is the applied indentation force, $n = ½$ and $c$ is the length from the center of the indentation to the end of the crack. Longer cracks thus result in lower fracture toughness values.

### A.3. *The Momentum Multiplication Factor ($\beta$)*

The momentum multiplication factor ($\beta$), necessary to connect the obtained mechanical properties with impact scenarios, is defined as the momentum change divided by the momentum input:

$$\beta = \frac{\Delta p_t}{M_p v_p} = \frac{M_p v_p + M_e v_e}{M_p v_p} = 1 + \frac{M_e v_e}{M_p v_p} \tag{6}$$

Here $\Delta p_t$ is the momentum change of the target due to the impact. $M_p$ and $v_p$ are the total mass and average velocity of the projectile, whereas $M_e$ and $v_e$ are the total mass and average velocity (on anti-impact direction) of the ejected material. If $\beta > 1$ there are fragments ejected after the impact, and the efficiency of an impact would be larger, since the attained $\Delta v$ would be also more significant. This effect is called "momentum multiplication" and has been studied by several authors (Hoerth et al. 2015).

The role of strength on impacts can be considered using the strength-dominated cratering model, which provides the following scaling relation (Holsapple & Housen 2012):

$$\beta - 1 \approx \left( v_p \sqrt{\frac{\rho_t}{Y}} \right)^{3\mu-1} \left( \frac{\rho_t}{\rho_p} \right)^{1-3\nu} \quad (7)$$

Where $v_p$ is the projectile velocity, $\rho_t$ and $\rho_p$ are the densities of the target and the projectile, $Y$ is some measure of the strength of the target, $\mu$ is a scaling parameter close to 0.55 for non-porous materials and between 1/3 and 0.4 for highly porous materials, and $\nu$ is a constant close to 0.4 for most target materials (Holsapple & Housen 2007). Many definitions of strength can be used for a geological material, but they all can be related on predictable ways (Holsapple 2009). We choose the compressive yield stress ($\sigma_C$) because it can be related to hardness following the typical expression $H = C\sigma_C$ in constrained materials (free from porosity), and because this relationship between hardness and compressive yield stress holds for the case of dynamic tests (Subhash et al. 1999). $C$ is the constraint factor and attains values close to 1.6 for rocks and 3 for metals, although values as high as 180 have been found for ceramics and other brittle materials (Zhang et al. 2011). Due to the size of nanoindentations, porosity, interactions between different phases, and cracks, are not perceived on these measurements. Altogether, those factors can imply a much higher $H$, and therefore $\sigma_C$, than expected for those materials at larger scales (Pellicer et al. 2012; Palchik & Hatzor 2004). Considering those points, we lack the means to calculate from our measurements of $H$ a value of $\sigma_C$ representative of a real asteroid, and therefore we cannot find β. However, we know that $H$ and $\sigma_C$ can be related through a specific constrain factor $C$. Therefore, for two similar materials A and B (and assuming the same $C$, density, and μ), we can calculate their ratio of β-1:

$$\frac{(\beta-1)_A}{(\beta-1)_B} \approx \left( \frac{Y_B}{Y_A} \right)^{\frac{3\mu-1}{2}} = \left( \frac{\sigma_{CB}}{\sigma_{CA}} \right)^{\frac{3\mu-1}{2}} = \left( \frac{H_B}{H_A} \right)^{\frac{3\mu-1}{2}} \quad (8)$$

That ratio provides us an idea of which of those materials would grant a higher momentum multiplication and a more efficient impact.